\documentclass[12pt]{article}

\usepackage{float}
\usepackage{amsfonts}

\usepackage{amssymb}
\usepackage{latexsym}
\usepackage{graphicx}
\usepackage[english]{babel}
\usepackage[font={small}]{caption}

\usepackage{amsfonts}
\usepackage{latexsym}
\usepackage{graphicx}
\usepackage[english]{babel}
\usepackage{tikz}

\begin{document}
\input epsf

\def\p{\partial}
\def\h{{1\over 2}}
\def\be{\begin{equation}}
\def\bea{\begin{eqnarray}}
\def\ee{\end{equation}}
\def\eea{\end{eqnarray}}
\def\d{\partial}
\def\la{\lambda}
\def\eps{\epsilon}
\def\bb{\bigskip}
\def\mm{\medskip}
\newcommand{\dm}{\begin{displaymath}}
\newcommand{\edm}{\end{displaymath}}
\renewcommand{\b}{\tilde{B}}
\newcommand{\gm}{\Gamma}
\newcommand{\ac}[2]{\ensuremath{\{ #1, #2 \}}}
\renewcommand{\ell}{l}
\newcommand{\z}{\ell}
\newcommand{\newsection}[1]{\section{#1} \setcounter{equation}{0}}
\def\bb{$\bullet$}
\def\Qbar{{\bar Q}_1}
\def\QPbar{{\bar Q}_p}

\def\q{\quad}

\def\bn{B_\circ}

\let\a=\alpha \let\b=\beta \let\g=\gamma \let\d=\delta \let\e=\epsilon
\let\c=\chi \let\th=\theta  \let\k=\kappa
\let\l=\lambda \let\m=\mu \let\n=\nu \let\x=\xi \let\r=\rho
\let\s=\sigma \let\t=\tau
\let\vp=\varphi \let\vep=\varepsilon
\let\w=\omega      \let\G=\Gamma \let\D=\Delta \let\Th=\Theta
                     \let\P=\Pi \let\S=\Sigma

\def\h{{1\over 2}}
\def\t{\tilde}
\def\r{\rightarrow}
\def\nn{\nonumber\\}
\let\bm=\bibitem
\def\Kt{{\tilde K}}
\def\b{\bigskip}

\let\p=\partial

\begin{flushright}
\end{flushright}
\vspace{20mm}
\begin{center}
{\LARGE The elastic vacuum\footnote{Essay awarded first prize in the Gravity Research Foundation 2021 Awards for Essays on Gravitation.}
 }
\\
\vspace{18mm}
 Samir D. Mathur

\vskip .1 in

 Department of Physics\\The Ohio State University\\ Columbus,
OH 43210, USA\\mathur.16@osu.edu\\
\vspace{4mm}
\end{center}
\vspace{10mm}
\thispagestyle{empty}
\begin{abstract}

The quantum gravity vacuum must contain virtual fluctuations of black hole microstates. These extended-size fluctuations get `crushed' when a closed trapped surface forms, and turn into on-shell `fuzzball' states that resolve the information puzzle. We argue that these same fluctuations can  get `stretched'  by the anti-trapped surfaces in an expanding cosmology, and that this stretching generates vacuum energy.  The stretching happen when the Hubble deceleration reduces quickly,  which happens whenever the pressure drops quickly. We thus get an inflation-scale vacuum energy when the heavy GUTS particles become nonrelativistic, and again a small vacuum energy  when the radiation phase turns to dust. 
The expansion law in the radiation phase does not allow stretching,  in agreement with the observed irrelevance of vacuum energy in that phase. The extra energy induced when the radiation phase changes to dust may explain the tension in the Hubble constant between low and high redshift data.

\end{abstract}
\vskip 1.0 true in

\newpage
\setcounter{page}{1}


In classical general relativity the Minkowski vacuum is empty and featureless. The cost of bending spacetime is captured by the curvature Lagrangian ${\mathcal R}$. Any cost for `stretching'  space would be captured by a cosmological constant $\Lambda$. But the values of $\Lambda$ suggested by observation are strange: an  energy density at the GUTS scale $\sim (10^{16}{\rm GeV})^4$ drives inflation, while  a vacuum energy  density  $\sim (10^{-3} {\rm eV})^4$ gives dark energy today. We will argue that vacuum energy at these disparate scales arises naturally from a nonperturbative feature of the quantum gravitational vacuum: the virtual fluctuations of black holes. 

In semiclassical gravitational collapse, matter in a star rushes to a central singularity, leaving  a vacuum around the horizon. Pair creation from this vacuum then leads to the information paradox \cite{hawking, cern}.  But in string theory the energy of the collapsing star goes to creating and exciting the extended objects in the theory -- strings and branes. This yields a horizon sized `fuzzball' \cite{emission}, which is essentially a `string star'. In geometric terms, the 6 compact dimensions cease to be trivially tensored with the 3+1 spacetime in the region $r\lesssim 2GM$. Pinching of compact circles creates KK-monopoles and antimopoles at various locations in this region. The monopole centers are linked by topological spheres, whose size is stabilized by gauge field fluxes \cite{fuzzballs}. There are a large number of such fuzzball configurations; these  yield the 
$
{\mathcal N}=Exp[{S_{bek}}]
$
 microstates of the hole, where $S_{bek}$ is the Bekenstein entropy \cite{bek}. 
 
\begin{figure}[H]
\begin{center}
\includegraphics[scale=.4]{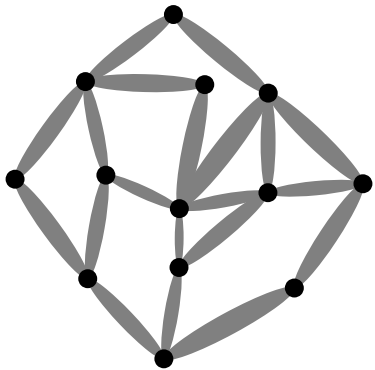}
\end{center}
\caption{A schematic depiction of a fuzzball. The dots represent KK monopoles or anti-monopoles, and the ellipses denote topologically $S^2$ surfaces between the centers of these monopoles and anti-monopoles. These spheres carry fluxes, which makes the structure resistant to compression and stretching.}
\label{fig9}
\end{figure}

 For our purposes below, we can model a fuzzball as a collection of masses (the monopoles) joined by springs (the spheres carrying fluxes); see Fig.\ref{fig9}.  The entire structure is a horizon-sized solution of nonperturbative string theory, having no horizon,  resistant to both compression and stretching. The fuzzballs radiate from their surface like normal bodies, so there is no information paradox.
 
If fuzzballs like Fig.\ref{fig9} exist as real objects for any mass $M$, then the vacuum must contain virtual fluctuations of these objects. The probability of such a fluctuation is estimated by $P=\left |A \right |^2$, with $A\sim Exp[-S_{grav}]$; here $S_{grav}$ is the gravitational action to create the configuration. Setting all length scales as order $\sim GM$, we find that
\be
S_{grav}\sim {1\over G}\int {\mathcal R}\sqrt{-g} \, d^4 x \sim  GM^2\sim \left ( {M\over m_p}\right)^2
\ee
As expected, $P\ll 1$ for $M\gg m_p$. But this smallness is offset by the very large {\it degeneracy} of fuzzball states of mass $M$ \cite{tunnel}
\be
{\mathcal N} \sim e^{S_{bek}}\sim e^{A\over 4G}\sim e^{4\pi GM^2}= e^{4\pi \left ( {M\over m_p}\right)^2}
\ee
{\it Thus the virtual fluctuations of  black hole microstates form an important component of the gravitational vacuum for all masses $0<M<\infty$.} We call these fluctuations {\it vecros}: ``virtual extended compression-resistant objects''. The compression-resistance of a vecro of radius $R_v$ is characterized by the energy increase $\Delta E$ when this radius is increased or decreased by a factor of order unity. The natural scales of the black hole suggest that
\be
\Delta E \sim M(R_v)
\label{six}
\ee
where $M(R_v)$ is the mass of a black hole of radius $R_v$. 

\begin{figure}[H]
\begin{center}
\includegraphics[scale=.55]{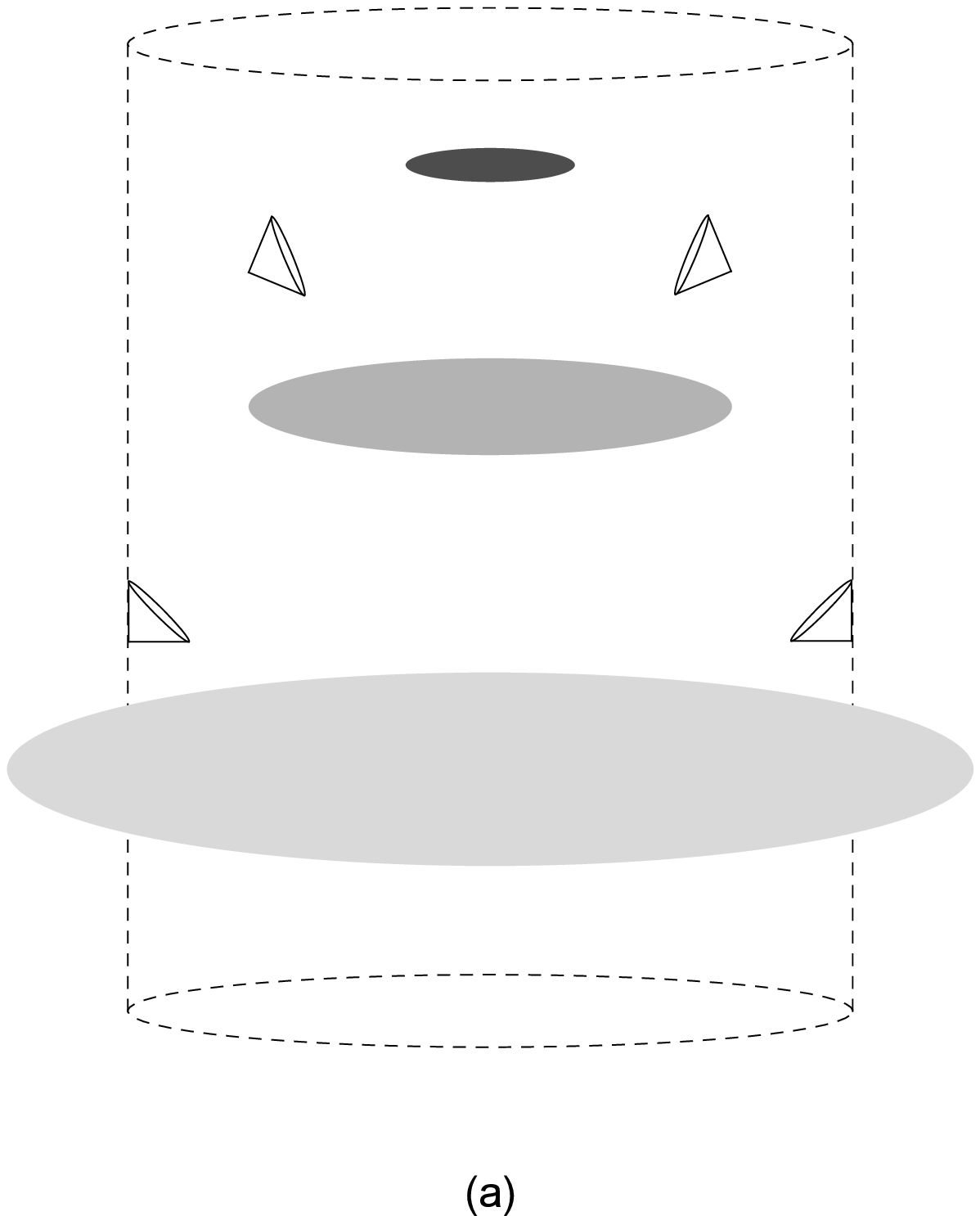}\hskip100pt
\includegraphics[scale=.55]{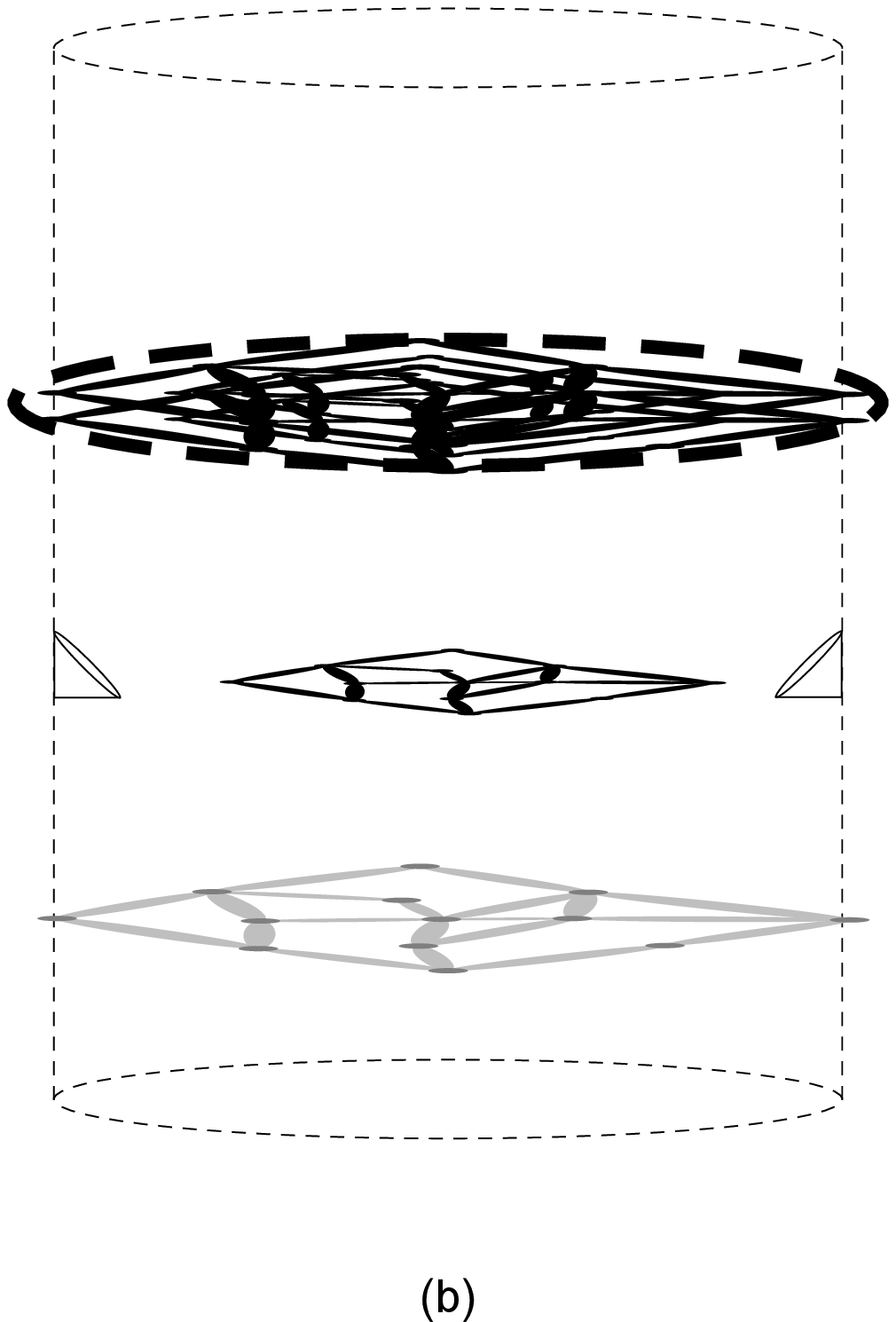}
\end{center}
\caption{(a) Semiclassical collapse: the gravitational collapse of a dust cloud generates a horizon; the light cones point inwards inside the horizon. (b) The modification due to vecros: at the bottom, a virtual fuzzball configuration (shown in grey); compression of this vecro leads to an on-shell excitation (shown higher up, in black); over a crossing timescale, such excitations generate a  horizon sized fuzzball (shown near the top).}
\label{fig20}
\end{figure}

What are the effects of vecros? Under small deformations of spacetime, the compression and stretching of the vecros is captured by the low energy Lagrangian ${\mathcal R}$. But consider the formation of a closed trapped surface in gravitational collapse.  
With semiclassical gravity, the equivalence principle holds near each point of the horizon $r_h=2GM$, so an   infalling shell passes uneventfully  through $r=r_h$.  Once the shell is inside the horizon, the inward-pointing  light cones prevent it from influencing pair creation near $r=r_h$; thus we cannot escape the information paradox caused by these pairs (Fig.\ref{fig20}a). But now consider  vecro structures in the vacuum wavefunctional, in the region $r<r_h$. Due to the inward-pointing nature of light cones, these vecros must keep compressing until their wavefunctional distorts by order unity (Fig.\ref{fig20}b). This distortion converts the virtual fluctuations to on-shell fuzzball states \cite{vecro}. The location $r\approx r_h$   thus changes from the vacuum to the surface of a fuzzball,  in a time of order crossing time $r_h$. As noted above,  the  fuzzball radiates like any normal body,  so we resolve the information paradox. Note that the extended and compression-resistant nature of the vecro fluctuations  allowed the vacuum to detect and react to the formation of a closed trapped surface,  thus bypassing the equivalence principle argument of semiclassical gravity. 

\begin{figure}[H]
\begin{center}
\includegraphics[scale=.60]{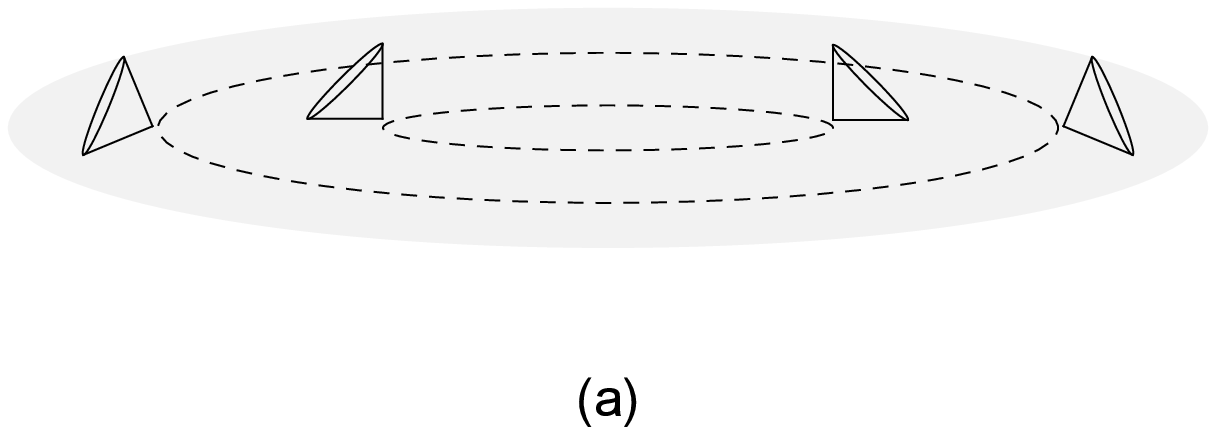}\hskip15pt
\includegraphics[scale=.60]{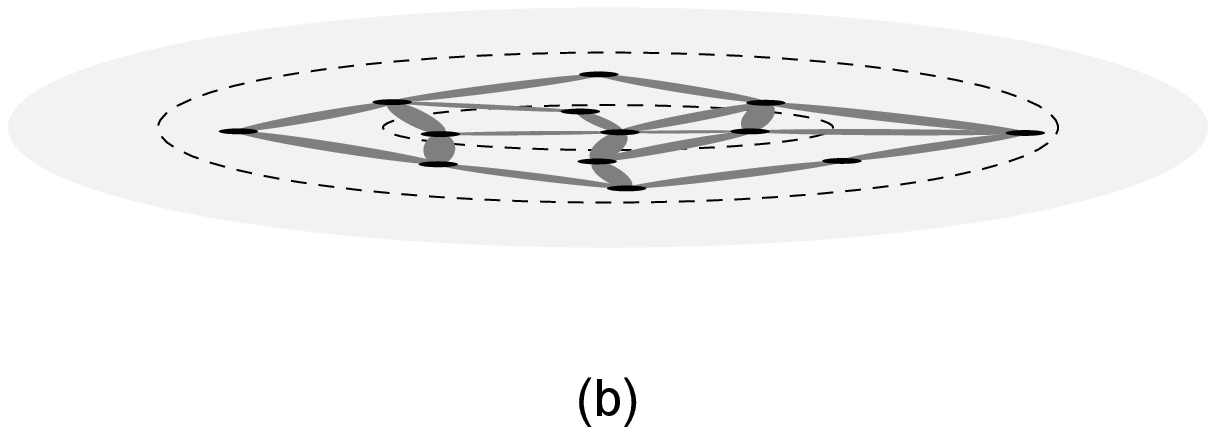}
\end{center}
\caption{(a) At the cosmological horizon $H^{-1}$ (inner dotted circle) the light cones are tangential to the surface of constant proper radius, and at larger radii the light cones point strictly outwards. (b) Vecro configurations larger than the cosmological horizon are forced to stretch due to the light cone structure.}
\label{fig3}
\end{figure}

An expanding cosmology is like a time-reversed black hole: we have anti-trapped surfaces where the light cones `point-outwards' (Fig.\ref{fig3}(a)). Consider a vecro structure that extends into this region having anti-trapped surfaces (Fig.\ref{fig3}(b)). The outward-pointing structure of light cones forces this vecro to stretch, just like the inward-pointing light cones in the black hole caused vecros to compress. This stretching of the vecros will yield the vacuum energy we need.

\b

\b

But will we not get this stretching energy at all times in the universe? Interestingly, the expansion of the universe makes it impossible to have vecro structures with arbitrarily large radii $R_v$. Consider a flat cosmology
\be
ds^2=-dt^2+a^2(t)[dr^2+r^2d\Omega_2^2]
\ee
Causality implies that vecro structures at any time $t$ cannot have a proper radius $R_v$ larger than the distance $R_{max}$ that light could have travelled since $t=0$. For the expansion
\be
a(t)=a_0 t^\alpha, ~~~H={\dot a \over a}={\alpha\over t}, ~~~H^{-1}={t\over \alpha}
\ee
we find that
\be
R_{max}(t)=a(t) \int_{t'=0}^t {dt'\over a(t')}={t\over 1-\alpha}
\ee
Thus as a fraction of the horizon radius $H^{-1}$, the size of the vecro fluctuations are limited to
\be
{R_v\over H^{-1}}\le {\alpha\over 1-\alpha}
\label{one}
\ee
In the radiation phase $\alpha=\h$, so we get
\be
{R_v\over H^{-1}}\le 1
\ee
so there are no vecros larger than the cosmological horizon. Thus there will be no stretching of vecros, and so no vacuum energy. This is consistent with  big bang nucleosynthesis, which disallows any significant source of energy in the radiation phase besides the standard model quanta.

In the dust phase $\alpha={2\over 3}$, so from (\ref{one})
\be
{R_v\over H^{-1}}\le { 2}
\ee
Now we {\it can} have vecros larger that the cosmological horizon; such vecros will have radii  $H^{-1}<R_v<{2}H^{-1}$. By the light cone structure depicted  in Fig.\ref{fig3}, these vecros will be forced to stretch and will therefore contribute a vacuum energy. 

What is the magnitude of this vacuum energy? The changeover from radiation domination to dust domination happens around $t=t_{equality}$ when the radiation and dust energy densities are equal.  From (\ref{six}), the scale of the stretching energy at this time is of order $M(H^{-1}_{equality})$, the mass of a black hole with radius of order the cosmological horizon. By the Hubble relation, such a mass corresponds to an energy density of order the closure density $\rho_{equality}$ at this time. Thus from  the stretched vecros we expect an energy density
\be
\rho_v\sim \alpha\,  \rho_{equality}
\ee
where $\alpha$ is a dimensionless constant. But what determines $\alpha$? If we adiabatically evolve the lowest energy state of a system, then it moves to a new lowest energy state.   Since there was no stretching energy in the radiation phase, adiabatic evolution would yield  a vanishing vacuum energy also in the dust phase;  i.e., we would get $\alpha=0$. If on the other hand we had a sudden transition from radiation to dust, then the vecro radius would keep growing for some time at the speed of light even in the dust phase, since this is the way it was growing in the radiation phase. Such an evolution would give $\alpha\sim 1$. In reality the transition from radiation to dust is neither adiabatic nor sudden; it happens over a few Hubble expansion times. Thus we expect $\alpha\ll1$ but not zero. This implies a vacuum energy which is a small fraction of the closure density at the time of matter-radiation equality, which is consistent with the observed value of dark energy today.

A somewhat similar change happens in the very early universe.  Above the GUTS scale, there is one species of radiation for each particle in the GUTS theory. Below the GUTS scale most of these particles become nonrelativistic, leaving only the standard model particles as  radiation. The Hubble expansion therefore suffers a change of the same type as in the radiation to dust transition. The stretching of vecros then gives a vacuum energy 
\be
\rho_v=\t \alpha\,  \rho_{GUTS}
\ee
where $\t \alpha$ is again a small dimensionless constant. Such a vacuum energy is consistent with the energy density needed for inflation. 

Recently it has been noted that there is a tension between the values of the Hubble constant predicted by low redshift and high redshift observations. Measurements on local objects suggest $H_0\approx 74 \, Km/s/Mpc$, while the $\Lambda$CDM model applied to cosmic microwave background measurements suggests $H_0\approx 67 \, Km/s/Mpc$. It has been noted that this tension can be resolved if there is `early dark energy': an extra vacuum energy around the time  of matter-radiation equality, with magnitude  $\sim 10\%$ of the closure density \cite{freese}. But as we have noted above, the stretching of vecros provides extra energy around just this time,  with magnitude a fraction of the closure density at that time. Thus the dynamics of vecros might resolve the tension in observed $H_0$ values. 

To summarize, the quantum gravitational vacuum must contain virtual fluctuations of black holes. The large degeneracy of black hole microstates makes these `vecro'  fluctuations an important component of the gravitational vacuum for all vecro radii $R_v$.  The inward pointing light cones inside a black hole horizon cause the  vecro configurations to crush, turning this part of the wavefunctional into on-shell fuzzballs; this resolves the information paradox.  The reverse situation presents itself in cosmology: we have anti-trapped surfaces outside the cosmological horizon, which force a stretching of vecros in that region to yield a vacuum energy. Interestingly, it appears that such an energy turns on just at the epochs where it is required for consistency with observations.

\b

\b

 \section*{Acknowledgements}

I would like to thank for useful discussions Robert Brandenberger, Sumit Das, Patrick Dasgupta,  Bin Guo,  Alan Guth, Emil Martinec, Anupam Mazumdar, Ken Olum and   Alex Vilenkin. This work is supported in part by DOE grant DE-SC0011726.

\b

\b


\end{document}